\documentclass[aps, twocolumn, showpacs, 10pt, superscriptaddress, preprintnumbers, nofootinbib]{revtex4-1}
\usepackage[colorlinks, citecolor=blue]{hyperref}
\usepackage{enumerate}
\usepackage{xcolor, graphicx}
\usepackage{latexsym, cancel, amssymb, amsmath, verbatim, mathrsfs}
\usepackage{mathtools}
\usepackage{physics, ulem}
\usepackage{subfigure}

\newcommand{\beq}{\begin{equation}}
\newcommand{\eeq}{\end{equation}}
\newcommand{\nn}{\nonumber}
\newcommand{\rmd}{\mathrm{d}}

\PassOptionsToPackage{normalem}{ulem}
\usepackage{ulem}
\RequirePackage{color}\definecolor{RED}{rgb}{1,0,0}\definecolor{BLUE}{rgb}{0,0,1} 

\allowdisplaybreaks[4]

\begin{document}

\title{Lam-Tung relation breaking effects and weak dipole moments at lepton colliders}

\author{Guanghui Li}
\email{ghli@ihep.ac.cn}
\affiliation{Institute of High Energy Physics, Chinese Academy of Sciences, Beijing 100049, China
}
\affiliation{School of Physical Sciences, University
of Chinese Academy of Sciences, Beijing 100049, China}
\author{Xu Li}
\email{xli@ge.infn.it (corresponding author)}
\affiliation{INFN, Sezione di Genova, Via Dodecaneso 33, I-16146 Genova, Italy
}
\author{Bin Yan}
\email{yanbin@ihep.ac.cn (corresponding author)}
\affiliation{Institute of High Energy Physics, Chinese Academy of Sciences, Beijing 100049, China
}
\affiliation{Center for High Energy Physics, Peking University, Beijing 100871, China}

\begin{abstract}
The breaking of the Lam-Tung relation in the Drell-Yan process at the LHC exhibits a long-standing tension with the Standard Model (SM) prediction at $\mathcal{O}(\alpha_s^3)$ accuracy. This tension could be explained by weak dipole interactions of leptons and quarks, associated with the $Z$-boson within the framework of the Standard Model Effective Field Theory (SMEFT). In this paper, we propose to cross-check these weak dipole interactions by measuring the violation effects of the Lam-Tung relation at future lepton colliders through the processes $e^+e^- \to Z\gamma \to \ell\bar{\ell}\gamma$ and $e^+e^- \to Z\gamma \to q\bar{q}\gamma$. By considering different decay modes of the $Z$-boson, these channels exhibit distinct sensitivities to various dipole operators, providing a way to disentangle their individual effects. Additionally, the high flavor-tagging efficiencies at lepton colliders could provide strong constraints on the dipole interactions of heavy quarks, such as $b$ and $c$ quarks, which are challenging to probe in the Drell-Yan process at  the LHC due to the suppression of parton distribution functions.
\end{abstract}

\maketitle

\section{Introduction}
The angular distributions of charged leptons in the Drell-Yan process~\cite{Drell:1970wh} at the Large Hadron Collider (LHC) provide valuable insights into the spin properties of the mediator $Z/\gamma^*$, the electroweak interactions between gauge bosons and fermions, and the dynamics of Quantum Chromodynamics (QCD). Within the Standard Model (SM), these distributions can be parameterized using a set of frame-dependent angular coefficients $A_i$ where $i=0,...,7$, which depend on the invariant mass, transverse momentum, and rapidity of the lepton pair. Notably, the angular coefficients $A_0$ and $A_2$ are known to satisfy the Lam-Tung relation~\cite{Lam:1980uc}, expressed as $A_0-A_2=0$. It has been demonstrated that this relation holds up to $\mathcal{O}(\alpha_s)$ but is violated at $\mathcal{O}(\alpha_s^2)$ and higher orders in perturbative QCD within the collinear factorization framework~\cite{Mirkes:1994dp,Karlberg:2014qua,Gauld:2017tww}. This violation can be interpreted as a consequence of the non-coplanarity of the hadron and parton planes at hadron colliders~\cite{Peng:2015spa}. The breaking of Lam-Tung relation has been experimentally verified by the ATLAS~\cite{ATLAS:2016rnf} and CMS~\cite{CMS:2015cyj} collaborations at $\sqrt{s} = 8\ \mathrm{TeV}$, as well as by the LHCb~\cite{LHCb:2022tbc} collaboration at $\sqrt{s}=13\ \mathrm{TeV}$. Their results reveal significant discrepancies from SM predictions at $\mathcal{O}(\alpha_s^2)$~\cite{Mirkes:1994dp,Karlberg:2014qua}, particularly in the high transverse momentum ($p_T$) region. These discrepancies could be significantly reduced when $\mathcal{O}(\alpha_s^3)$~\cite{Gauld:2017tww} correction are  considered, as the leading-order effects contributing to this violation originate at $\mathcal{O}(\alpha_s^2)$. However, deviation persist for $p_T>50~{\rm GeV}$ in the ATLAS measurements when compared to the regularized data with the $\mathcal{O}(\alpha_s^3)$ prediction. Electroweak corrections to $A_0-A_2$ have also been discussed in Ref.~\cite{Frederix:2020nyw}, but they fail to explain the observed anomalies as well.  Although non-perturbative effects beyond the collinear factorization, such as Boer-Mulders functions, could contribute to this observable, their effects are more pronounced in the lower $p_T$ region~\cite{Zhou:2009rp,Nefedov:2020ugj,Balitsky:2021fer,Piloneta:2024aac}, which lies far from the region where discrepancies are observed. 

Although it is too early to claim that this discrepancy originates from new physics (NP) beyond the SM, it is still interesting to explore this possibility, as $A_0-A_2$ can be measured with very high precision at the high-luminosity LHC  (HL-LHC). In a recent paper~\cite{Li:2024iyj}, we demonstrated that the leading contribution to this observable arises from the quadratic effects of dimension-6 dipole operators associated with the $Z$ boson, calculated with $\mathcal{O}(\alpha_s)$ accuracy in QCD within the framework of the Standard Model Effective Field Theory (SMEFT) and could provide a compelling explanation for the observed discrepancy. However, the hadronic process $pp \to ZX \to \ell\bar{\ell} X$ involves interactions of the $Z$-boson with both quarks and leptons, making it challenging to directly distinguish their contributions solely based on the $A_0-A_2$ measurement. It is well known that the investigation of dipole moments is essential for exploring the intrinsic quantum properties of particles, and further studies in this direction could shed light on the underlying mechanisms responsible for the observed tension. Recently, several approaches leveraging the transverse spin information of quarks and leptons have been proposed to test these dipole interactions and demonstrated that these novel spin observables have the potential to enhance current constraints by one to two orders of magnitude, while remaining free from contamination of other NP effects~\cite{Boughezal:2023ooo,Wen:2023xxc,Shao:2023bga,Wang:2024zns,Wen:2024cfu,Wen:2024nff}.

\begin{figure}[tb]
    \centering
    \includegraphics[scale=0.5]{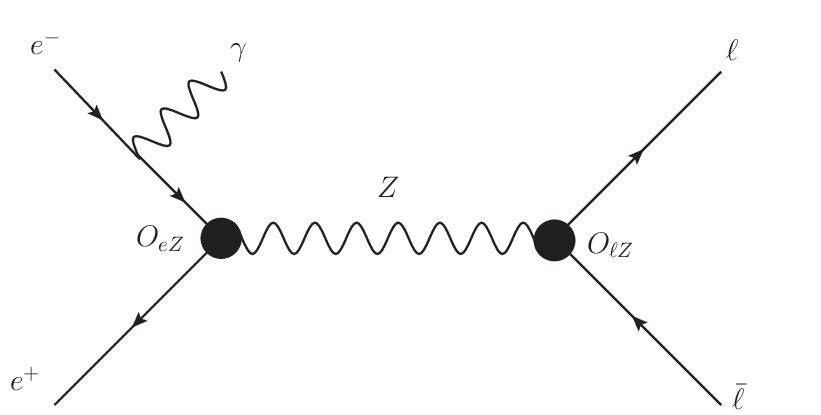}
\caption{Illustrative Feynman diagram for the $e^+ e^-\to Z\gamma\to \ell\bar{\ell}\gamma$ production at the lepton colliders. The black dots denote the effective dipole couplings from the $Z$-boson production ($O_{eZ}$) and decay ($O_{\ell Z}$).}
\label{fig:123}
\end{figure}

In this work, we propose to study the breaking effects of Lam-Tung relation at future lepton colliders, such as the Circular Electron-Positron Collider (CEPC)~\cite{CEPCStudyGroup:2018ghi,CEPCStudyGroup:2023quu}, International Linear Collider (ILC)~\cite{ILC:2013jhg}, and the Future Circular Collider (FCC-ee)~\cite{FCC:2018evy}. We focus on the processes $e^+e^-\to Z\gamma \to \ell\bar{\ell} \gamma$ and $e^+e^-\to Z\gamma \to q\bar{q} \gamma$, as illustrated in Fig.~\ref{fig:123}, to address the degeneracy of the dipole operators associated with quarks and leptons observed in LHC $A_0-A_2$ measurements~\cite{Li:2024iyj}. Since these processes involve different final-state particles, they are sensitive to distinct dipole operators, allowing us to disentangle their individual contributions.

\section{Lam-Tung relation breaking in the SMEFT}

The angular distribution of leptons in the process $e^+ e^- \to Z\gamma X \to\ell \bar{\ell} \gamma X$ can be expanded in terms of  harmonic polynomials with $l\leq 2$, owing to the spin-1 nature of the $Z$-boson in the SM~\cite{Collins:1977iv}, 
\beq
\begin{aligned}
&\frac{\rmd\sigma}{\rmd p_T \rmd m_{\ell\bar{\ell}}\rmd \Omega}=\frac{3}{16\pi}\frac{\rmd\sigma}{\rmd p_T \rmd m_{\ell\bar{\ell}}}\times \bigg\{(1+\cos^2\theta)\\
&+\frac{1}{2}A_0(1-3\cos^2\theta)+A_1\sin2\theta\cos\phi+\frac{1}{2}A_2\sin^2\theta\cos2\phi\\
&+A_3\sin\theta\cos\phi+A_4\cos\theta+A_5\sin^2\theta\sin2\phi\\
&+A_6\sin2\theta\sin\phi+A_7\sin\theta\sin\phi\bigg\}, 
\end{aligned}
\label{AngExp}\eeq
where $\theta$ and $\phi$ represent the polar and azimuthal angles of the lepton in the Collins-Soper frame~\cite{Collins:1977iv}, and $p_T$ and $m_{\ell\bar{\ell}}$ denote the transverse momentum and the invariant mass of the lepton pair, respectively. The angular coefficients $A_i$ can be extracted from the differential cross section by taking moments of their corresponding angular distribution functions, leveraging the orthogonality of harmonic polynomials,
\beq
\begin{aligned}
A_0 &= 4-10\langle \cos^2\theta \rangle, & A_1&=5\langle\sin2\theta\cos\phi\rangle, \\
A_2&=10\langle\sin^2\theta\cos2\phi\rangle, & A_3&=4\langle \sin\theta\cos\phi\rangle, \\
A_4&=4\langle\cos\theta\rangle, & A_5 &= 5 \langle \sin^2\theta\sin 2\phi\rangle, \\
A_6 &= 5\langle\sin2\theta\sin\phi\rangle, & A_7 &= 4 \langle \sin\theta\sin\phi\rangle, 
\end{aligned}
\nn\eeq
where, 
\beq
\langle \omega(\theta, \phi) \rangle=\frac{1}{\frac{\rmd\sigma}{\rmd p_T \rmd m_{\ell\bar{\ell}}}}\int\rmd\Omega\frac{\rmd\sigma}{\rmd p_T \rmd m_{\ell\bar{\ell}}\rmd\Omega}\omega(\theta, \phi).
\label{moment}
\eeq
It is interesting to consider all possible angular coefficients $A_i$ at lepton colliders to constrain NP beyond the SM, while this study will focus specifically on the $A_0 - A_2$ measurement and defer the impact of others to future work. This choice is motivated by the fact that $A_0 - A_2$ has been demonstrated to be particularly sensitive to dipole interaction of quarks and leptons~\cite{Li:2024iyj}.

Within the framework of the SMEFT~\cite{Buchmuller:1985jz,Grzadkowski:2010es}, these dipole interactions can be parameterized as,
\beq
\begin{aligned}
\mathcal{L}_{\psi^2 X \varphi}=
&\bar{q}_L\sigma^{\mu\nu}(C_{uB}B_{\mu\nu}+C_{uW}\tau^IW^I_{\mu\nu})\frac{\tilde{\varphi}}{\Lambda^2}u_R\\
+&\bar{q}_L\sigma^{\mu\nu}(C_{dB}B_{\mu\nu}+C_{dW}\tau^IW^I_{\mu\nu})\frac{\varphi}{\Lambda^2}d_R\\
+&\bar{\ell}_L\sigma^{\mu\nu}(C_{eB}B_{\mu\nu}+C_{eW}\tau^IW^I_{\mu\nu})\frac{\varphi}{\Lambda^2}e_R+h.c.\ , 
\end{aligned}
\eeq
where $q_L(\ell_L), u_R, d_R$, and $e_R$ represent the left-handed quark (lepton) doublet, right-handed up-type quark, down-type quark, and lepton fields respectively. $B_{\mu\nu}$ and $W^I_{\mu\nu}$ denote the field strength tensor of $U(1)_Y$ and $SU(2)_L$ gauge fields, and $\varphi$ represents the Higgs doublet field. The $C_i$ and $\Lambda$ are corresponding Wilson coefficient and scale of the NP. After the electroweak symmetry breaking, $\langle \varphi \rangle = v/\sqrt{2}$ with $v=246~{\rm GeV}$ , the dipole operators relevant to the process $e^+e^-\to Z\gamma \to f\bar{f}\gamma$, with $f=u,d,\ell$ can be expressed as,
\beq
\begin{aligned}
O_{f Z}=&\frac{v}{\sqrt{2}\Lambda^2}\bar{f}_L \sigma^{\mu\nu}f_R Z_{\mu\nu}, &
O_{f A}=&\frac{v}{\sqrt{2}\Lambda^2}\bar{f}_L \sigma^{\mu\nu}f_R A_{\mu\nu}.
\end{aligned}
\nn\eeq
The Wilson coefficients for the above operators are given by 
\beq
\begin{aligned}
C_{\ell A}&=c_W C_{\ell B} - s_W C_{\ell W}, & C_{\ell Z}&=-s_W C_{\ell B} - c_W C_{\ell W},\nn\\
C_{dA}&=c_W C_{dB} - s_W C_{dW}, & C_{dZ}&=-s_W C_{dB} - c_W C_{dW}, \nn\\
C_{uA}&=c_W C_{uB} + s_W C_{uW}, & C_{uZ}&=-s_W C_{uB} + c_W C_{uW}, \nn   
\end{aligned}
\eeq
where $c_W (s_W)$ represents the cosine (sine) of the weak mixing angle. By focusing on the invariant mass region around the $Z$-pole, contributions from the photon can be neglected. Additionally, the coefficients $C_{fA}$ are also strongly constrained by measurements of the anomalous magnetic and electric dipole moments of fermions~\cite{Pitschmann:2014jxa,Liu:2017olr,Aebischer:2021uvt,ParticleDataGroup:2024cfk}. 

\begin{figure}[tb]
    \centering
    \includegraphics{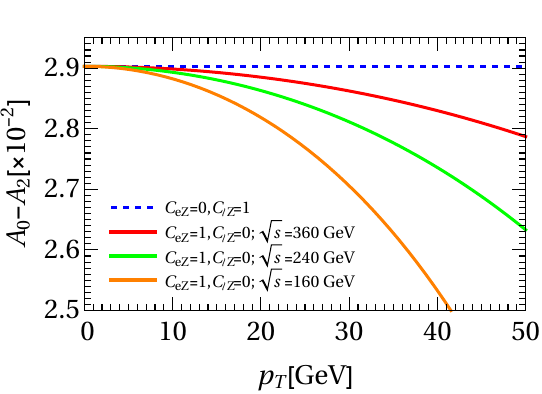}
\caption{The contribution of dipole operators to the $A_0-A_2$ as a function of $p_T$ for the different lepton collider energies, assuming $C_{eZ}=1$ or $C_{\ell Z}=1$ and $\Lambda=1~{\rm TeV}$.}
\label{fig:plot}
\end{figure}

Next, we analyze the angular coefficients $A_0$ and $A_2$  in the process $e^+e^- \to Z\gamma\to \ell \bar{\ell} \gamma$, as illustrated in Fig.~\ref{fig:123}. This analysis will subsequently be extended to include quark final states. It is worth noting that the quark and anti-quark can generate the same coefficients $A_0$ and $A_2$ in Eq.~\ref{AngExp}, as a result, we do not distinguish between them in the analysis. Neglecting the lepton masses, the analytical expressions of $A_0$ and $A_2$ are given by:
\beq 
\begin{aligned}
(A_0)_\ell &= \frac{p_T^2}{m_{\ell\bar{\ell}}^2+p_T^2}\\ 
&+\frac{C_{eZ}^2}{\Lambda^4}\frac{4v^2 \left[(m_{\ell\bar{\ell}}^4-s p_T^2)^2+s^2 (m_{\ell\bar{\ell}}^2+p_T^2)^2 \right]}{(g_L^2+g_R^2)(m_{\ell\bar{\ell}}^2+p_T^2)(m_{\ell\bar{\ell}}^4 -2sp_T^2+s^2)} \\
&+\frac{C_{\ell Z}^2}{\Lambda^4}\frac{4v^2  m_{\ell\bar{\ell}}^4  }{(g_L^2+g_R^2)(m_{\ell\bar{\ell}}^2+p_T^2)},\\
(A_2)_\ell &= \frac{p_T^2}{m_{\ell\bar{\ell}}^2+p_T^2}\\ 
&+\frac{C_{eZ}^2}{\Lambda^4}\frac{4v^2 p_T^2\left[-m_{\ell\bar{\ell}}^6+2m_{\ell\bar{\ell}}^4s+m_{\ell\bar{\ell}}^2s(4p_T^2+s)+2p_T^2s^2 \right]}{(g_L^2+g_R^2)(m_{\ell\bar{\ell}}^2+p_T^2)(m_{\ell\bar{\ell}}^4 -2sp_T^2+s^2)} \\
&-\frac{C_{\ell Z}^2}{\Lambda^4}\frac{4v^2 p_T^2 m_{\ell\bar{\ell}}^2  }{(g_L^2+g_R^2)(m_{\ell\bar{\ell}}^2+p_T^2)},
\label{eq:A0A2}
\end{aligned}
\eeq
where $g_L=g/c_W(-1/2+s_W^2)$ and $g_R=g/c_Ws_W^2$ represent the left-handed and right-handed couplings between the $Z$-boson and leptons in the SM. With the results presented in Eq.~\ref{eq:A0A2}, several comments are in order:
\begin{itemize}
\item Since dipole interactions flip fermion helicities, their interference with the SM amplitudes at $\mathcal{O}(1/\Lambda^2)$ vanishes for light fermions due to the chiral symmetry. Consequently, the leading contribution to the $A_0-A_2$ arises at $\mathcal{O}(1/\Lambda^4)$.
\item In the limit of $p_T\to 0$, the angular coefficient $(A_2)_\ell\to 0$, while $(A_0)_\ell$ remains non-zero when considering dipole interactions. This behavior is consistent with the process $e^+e^-\to Z\to \ell\bar{\ell}$, where the differential distribution becomes independent of the $\phi$ angle, as the $Z$-boson cannot be linearly polarized.
\item In the limit of $p_T\to 0$, we find identical coefficients for $C_{eZ}$ and $C_{\ell Z}$ in the  angular coefficient $(A_0)_\ell$. This can be understood as a consequence of $C_{eZ}$ and $C_{\ell Z}$ corresponding to dipole interactions in the production and decay of the $Z$-boson, respectively. Under the narrow width approximation, the cross section of $e^+e^-\to Z\to \ell\bar{\ell}$ is proportional to $\Gamma(Z\to e^+e^-)\Gamma(Z\to \ell\bar{\ell})$, where $\Gamma$  denotes the partial decay width of the $Z$ boson.
\item As expected, $A_0=A_2$ in the SM at leading order (LO) for $Z+\gamma$ production.
\end{itemize}
Therefore, we derive the contributions of dipole interactions to the breaking effects of the Lam-Tung relation,
\beq
\begin{aligned}
(A_0-A_2)_\ell=&\frac{4v^2 m_{\ell\bar{\ell}}^2  }{(g_L^2+g_R^2)}\times \\ &\left[\frac{C_{eZ}^2}{\Lambda^4}\left(1-\frac{2s p_T^2}{m_{\ell\bar{\ell}}^4 -2s p_T^2+s^2}\right) +\frac{C_{\ell Z}^2}{\Lambda^4}\right].
\end{aligned}
\eeq
Figure~\ref{fig:plot} shows the predicted $A_0-A_2$ from dipole interactions involving the $Z$ boson, with $C_{eZ} = 1$ or $C_{\ell Z}=1$ and $\Lambda = 1~\text{TeV}$, for the process $e^+e^- \to Z\gamma \to \ell\bar{\ell}\gamma$ at a dilepton invariant mass $m_{\ell\bar{\ell}} = m_Z$. It clearly demonstrates that the contribution from the dipole interaction in $Z$-boson decay ($C_{\ell Z}$) is independent of $p_T$ and the collider energy $\sqrt{s}$ due to the cancellation between $A_0$ and $A_2$, while the contribution from the production ($C_{eZ}$) remains sensitive to $p_T$ and $\sqrt{s}$. This distinction makes it possible to identify the source of the dipole interactions from $A_0-A_2$ distribution.   
The process $e^+e^- \to Z\gamma \to q\bar{q}\gamma$ yields the same expression for $A_0 - A_2$, differing only in the coupling constants $g_L$ and $g_R$ for the $C_{qZ}$ term, which should be adjusted to correspond to the specific quark flavor, i.e,
\beq
\begin{aligned}
(A_0-A_2)_q=&\frac{4v^2 m_{q\bar{q}}^2  }{(g_L^2+g_R^2)}\frac{C_{eZ}^2}{\Lambda^4}\left(1-\frac{2s p_T^2}{m_{\ell\bar{\ell}}^4 -2s p_T^2+s^2}\right)+  \\ &\frac{4v^2 m_{q\bar{q}}^2  }{(g_{L,q}^2+g_{R,q}^2)}\frac{C_{q Z}^2}{\Lambda^4},
\end{aligned}
\eeq
where $g_{L,q}=g/c_W(T_3^q-Q_qs_W^2)$ and $g_{R,q}=-g/c_W Q_qs_W^2$ represent the left-handed and right-handed coupling between the $Z$-boson and quark $q$ in the SM. However, the additional quark flavor separation would strongly depend on the jet charge and flavor tagging efficiencies~\cite{Field:1977fa,Krohn:2012fg,Waalewijn:2012sv,Fraser:2018ieu,Li:2019dre,Kang:2020fka,Kang:2021ryr,Kang:2023ptt,Wang:2023azz,Li:2023tcr,Cui:2023kqb}, which are crucial for probing the flavor dependence of the quark dipole interactions.

\section{Numerical Results}

\begin{figure*}[tb]
    \centering
    \includegraphics[scale=1.2]{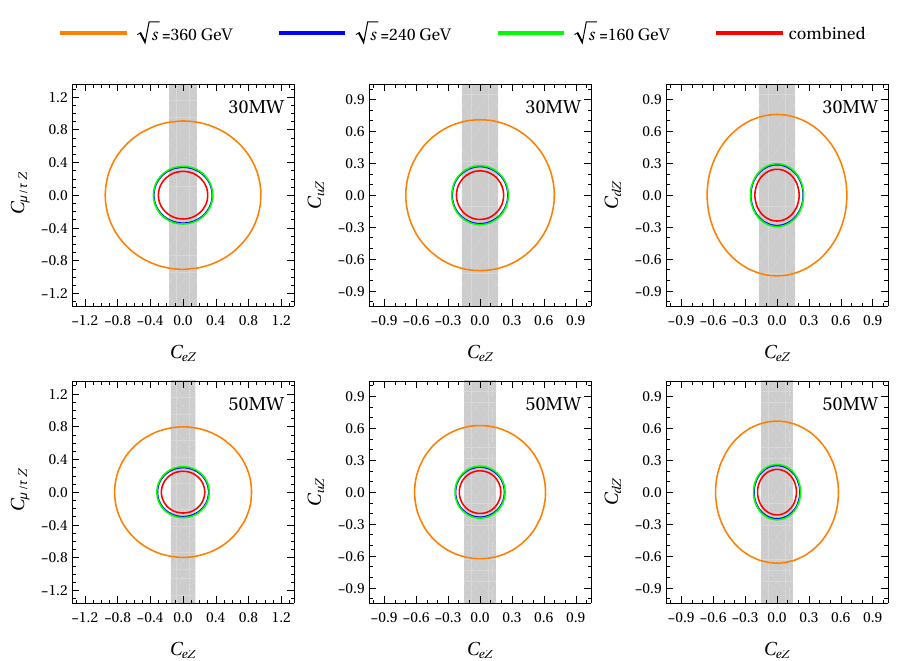}
\caption{Expected constraints on the weak dipole interactions of quarks and leptons at future lepton colliders, obtained from $A_0-A_2$ measurements in the processes $e^+e^- \to Z\gamma \to \ell\bar{\ell}\gamma$ and $e^+e^- \to Z\gamma \to q\bar{q}\gamma$. The decay channel $Z \to e^+e^-$ is exclusively sensitive to $C_{eZ}$, with combined results represented by gray regions. The constraints are shown at 68\% C.L. for different beam radiation power scenarios.}
\label{fig:chi2}
\end{figure*}

We now present the projected constraints on the dipole interactions from the $A_0 - A_2$ measurements at the CEPC. The CEPC is expected to operate at four distinct center-of-mass energy levels: 240 GeV for Higgs boson studies, 160 GeV for $W$ boson investigations, 91 GeV for $Z$ boson studies, and 360 GeV for top quark physics. Although $\sqrt{s} = 91~\rm{GeV}$ could also be used to study dipole interactions, it could be demonstrated that $e^+e^- \to Z \to f\bar{f}\gamma$ can generate a violation of the Lam-Tung relation at LO in the SM. As a result, many potential NP effects could contaminate the measurements of dipole interactions, and this scenario will not be considered in this study. 

To avoid the impact of infrared divergences from soft and collinear photons, we require $p_T > 2~\rm{GeV}$ in our analysis. The kinematic cut on the invariant mass of the final states from $Z$ boson decay depends on the mass resolution in lepton colliders. It has been shown that a per mille-level resolution can be achieved for leptonic final states, while it is approximately 4\% for hadronic states at the CEPC~\cite{CEPCStudyGroup:2018ghi,CEPCStudyGroup:2023quu}. For a conservative estimation, we adopt a 4\% mass resolution for all final states and the conclusions are not sensitive to this choice.  After integrating over the kinematic regions of $p_T$ and the invariant mass of the final states, we obtain the dependence of $\Delta A=A_0-A_2$ on the Wilson coefficient $C_{fZ}$ for different collider energy $\sqrt{s}$. 

For $\sqrt{s}=160~\rm GeV$, we have 
\beq
\begin{aligned}
\Delta A_\ell&=0.0275 \ C_{eZ}^2+0.0291 \ C_{\ell Z}^2,\\
\Delta A_{u}&=0.0275 \ C_{eZ}^2+0.0255 \ C_{u Z}^2,\\
\Delta A_{d}&=0.0275 \ C_{eZ}^2+0.0198 \ C_{d Z}^2,
\end{aligned}
\eeq
where $u,d$ represent the up and down type quarks respectively. 

For $\sqrt{s}=240~\rm GeV$, we have 
\beq
\begin{aligned}
\Delta A_\ell&=0.0266 \ C_{eZ}^2+0.0291 \ C_{\ell Z}^2,\\
\Delta A_{u}&=0.0266 \ C_{eZ}^2+0.0255 \ C_{u Z}^2,\\
\Delta A_{d}&=0.0266 \ C_{eZ}^2+0.0198 \ C_{d Z}^2.
\end{aligned}
\eeq

For $\sqrt{s}=360~\rm GeV$, we have 
\beq
\begin{aligned}
\Delta A_\ell&=0.0264 \ C_{eZ}^2+0.0291 \ C_{\ell Z}^2,\\
\Delta A_{u}&=0.0264 \ C_{eZ}^2+0.0255 \ C_{u Z}^2,\\
\Delta A_{d}&=0.0264 \ C_{eZ}^2+0.0198 \ C_{d Z}^2.
\end{aligned}
\eeq

To estimate the expected sensitivities for probing dipole interactions, we conduct a $\chi^2$ analysis as follows,
\beq
\chi^2 = \left[ \frac{\Delta A_{\rm exp}-\Delta A_{\rm {SMEFT}}-\Delta A_{\rm {SM}}}{\delta\Delta A} \right]^2, 
\eeq
where $\Delta A_{\rm exp}$, $\Delta A_{\rm SMEFT}$, and $\Delta A_{\rm SM}$ represent the experimental measurements at the CEPC, the contributions from dipole operators, and the SM prediction, respectively. For simplicity, we have assumed that the experimental values are consistent with the SM predictions. The term $\delta\Delta A$ denotes the corresponding statistical uncertainty of $A_0-A_2$~\footnote{Systematic uncertainties could arise from various experimental effects, such as cut acceptance, tagging efficiency, detector resolution, and luminosity. For simplicity, we will ignore these effects in this study.}, whose value can be calculated as
\beq
\begin{aligned}
[\delta\Delta A]^2&\equiv \left[\delta\left(\frac{\sum^N_{i=1} \omega(\theta_i, \phi_i)}{N}\right)\right]^2\\
&=\sum_{i=1}^N[\delta\omega(\theta_i, \phi_i)]^2\times \left(\frac{1}{N}\right)^2 \\
&=\sum_{i=1}^N \int \frac{\mathrm{d}\sigma}{\sigma}\omega^2(\theta_i,\phi_i)\times \left(\frac{1}{N}\right)^2\\
&=\frac{228}{7}\times \frac{1}{N},
\end{aligned}
\eeq
where $\omega(\theta,
\phi)=4-10\cos^2\theta-10\sin^2\theta\cos2\phi$, $N$ represents the event number with $N=\sigma \mathcal{L}$. Here, we focus on the LO cross section within the SM, and the relation $A_0=A_2$ has been utilized in above uncertainty estimation. The expected integrated luminosity depends on the single beam radiation power and the collider energies. We summarize the expected luminosity and the statistical error $\delta\Delta A$ at the CEPC for different collider energies in Table~\ref{tab:CEPC}.
\begin{table}[htbp]
\centering
\begin{tabular}{c|c|c|c}
\hline
 $\sqrt{s}~({\rm GeV})$& 160 & 240 & 360 \\
\hline
30 MW, $\mathcal{L}~({\rm ab}^{-1})$ & 4.2  & 13 & 0.6\\
\hline
30 MW, $\delta\Delta A_\ell~(\times 10^{-3})$ & 2.4& 2.2&16\\
\hline
30 MW, $\delta\Delta A_u~(\times 10^{-3})$ & 1.3& 1.2&8.5\\
\hline
30 MW, $\delta\Delta A_d~(\times 10^{-3})$ & 1.1& 1.1&7.5\\
\hline
50 MW, $\mathcal{L}~({\rm ab}^{-1})$ & 6.9 & 22 & 1.0\\
\hline
50 MW, $\delta\Delta A_\ell~(\times 10^{-3})$& 1.9& 1.7& 12 \\
\hline
50 MW, $\delta\Delta A_u~(\times 10^{-3})$& 1.0& 0.9& 6.6 \\
\hline
50 MW, $\delta\Delta A_d~(\times 10^{-3})$& 0.9& 0.8& 5.8 \\
\hline
\end{tabular}
\caption{The expected luminosity and statistical error $\delta\Delta A$ at different collider energies and beam radiation powers at the CEPC~\cite{CEPCStudyGroup:2018ghi,CEPCStudyGroup:2023quu}.}\label{tab:CEPC}
\end{table}
By requiring $\chi^2<2.3$, we obtain the expected limits for the dipole couplings at the $68\%$ confidence level (C.L.) for different collider energies and processes, as shown in Fig.~\ref{fig:chi2}. All the dipole couplings could be constrained to be of $\mathcal{O}(0.1)$. Notably, these results can also be applied to the heavy-flavor dipole couplings (e.g., $b$ and $c$ quarks) after imposing flavor tagging requirements. The results are anticipated to remain largely unchanged, owing to the high flavor-tagging efficiencies at the CEPC~\cite{CEPCStudyGroup:2018ghi,CEPCStudyGroup:2023quu}, which are approximately 80\% for $b$ quarks and 60\% for $c$ quarks. If we focus on the decay channel $Z \to e^+e^-$, the observable $\Delta A$ becomes solely dependent on the electron dipole coupling $C_{eZ}$. We present the combined results for this scenario in Fig.~\ref{fig:chi2}, indicated by the gray region. This approach enables us to isolate the contribution of the electron dipole interaction while achieving a clear separation of the various contributions arising from different dipole interactions.

\begin{figure}[tb]
    \centering
    \includegraphics[scale=0.47]{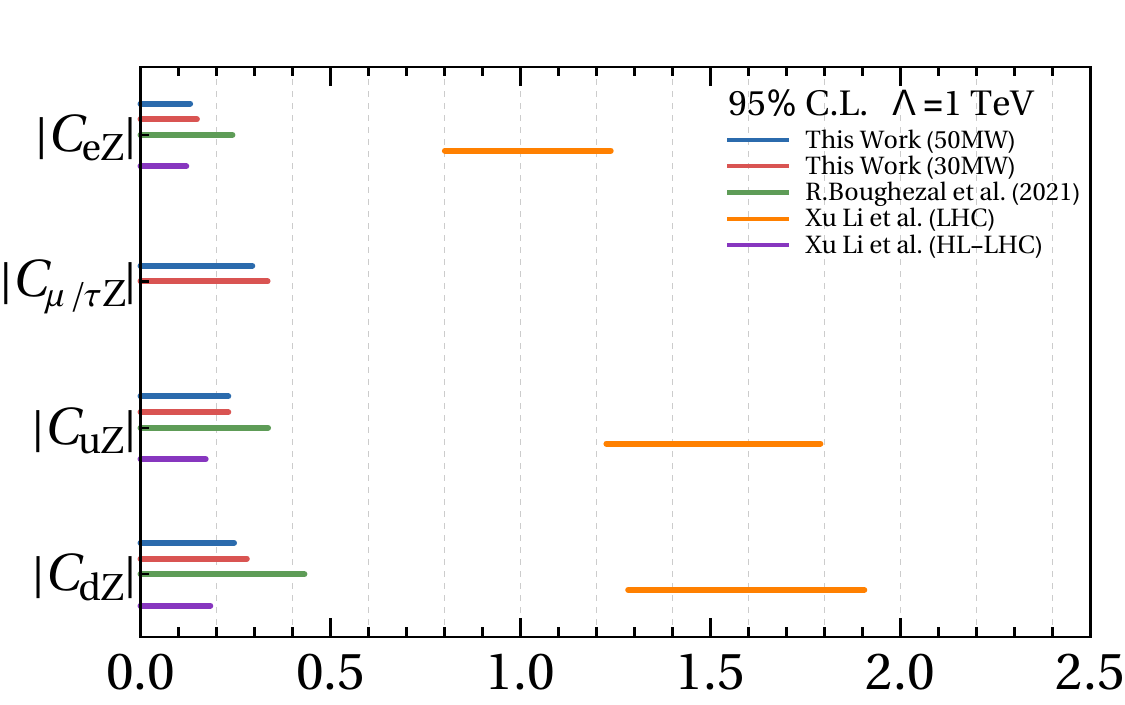}
\caption{Combined individual bounds on the Wilson coefficients of weak dipole operators at 95\% C.L. with $\Lambda=1~{\rm TeV}$ for different beam radiation powers at the CEPC (blue and red lines). The green and orange lines correspond to the results from the invariant mass~\cite{Boughezal:2021tih} and $A_0-A_2$~\cite{Li:2024iyj} distributions, respectively, in Drell-Yan production at the $\sqrt{s}=8~{\rm TeV}$ LHC. The purple lines indicate the projected sensitivity from the $A_0-A_2$ distributions at the HL-LHC~\cite{Li:2024iyj}.} 
\label{fig:bounds}
\end{figure}

We derive combined individual bounds for each operator by considering one operator at a time at a 95\% C.L., as illustrated in Fig.~\ref{fig:bounds}. For the electron dipole coupling $C_{eZ}$, we consider all possible decay modes of the $Z$ boson, excluding the neutrino final states.
These results are compared with those obtained from the analysis of existing LHC data~\cite{daSilvaAlmeida:2019cbr,Boughezal:2021tih,Li:2024iyj, Gauld:2024glt, Hiller:2025hpf}. The significant deviation in $C_i$ (orange lines in Fig.~\ref{fig:bounds}) reported in Ref.~\cite{Li:2024iyj} stems from the tension in the $A_0-A_2$ measurements observed by the ATLAS collaboration.  Although the expected limits from $A_0-A_2$ measurements at the CEPC are comparable to those from the LHC, several advantages of the CEPC approach are worth emphasizing, as outlined below:
\begin{itemize}
    \item The dipole interactions can contribute to the $A_0-A_2$ as a leading effect, while they play a subleading role in other observables, such as the invariant mass and transverse momentum of leptons~\cite{Boughezal:2021tih}. Consequently, the conclusions in Ref.~\cite{Boughezal:2021tih} strongly depend on the theoretical assumptions.
    \item The degeneracy between quark and lepton dipole interactions in Ref.~\cite{Li:2024iyj} can be resolved by analyzing different $Z$ boson decay modes at the CEPC.
    \item For heavy flavor quarks, such as $b$ and $c$ quarks, the results are suppressed by the parton distribution functions in the Drell-Yan process at the LHC. In contrast, the limits derived in our study are nearly identical for both heavy and light quarks due to the high flavor-tagging efficiencies.
\end{itemize}

\section{Conclusions}
In this paper, we propose to use the breaking effects of the Lam-Tung relation in the scattering process $ e^+e^- \to Z\gamma \to \ell\bar{\ell}\gamma/ q\bar{q}\gamma $ at future lepton colliders to investigate potential new physics effects arising from weak dipole interactions between the $Z$ boson and quarks/leptons. Compared to the Drell-Yan process at hadron colliders, these violation effects are only subject to perturbative corrections and are free from contamination by non-perturbative effects, making them a promising smoking-gun signal of weak dipole interactions. By employing jet charge and flavor tagging techniques, it is possible to distinguish between different decay modes of the $Z$ boson. This capability enables the unambiguous separation of dipole operators, a feat unattainable in hadronic experiments.

\vspace{3mm}
\noindent{\bf Acknowledgments.}
G. Li and B.Yan were supported in part by the National Science Foundation of China under Grant No.~12422506, the IHEP under Grant No.~E25153U1 and CAS under Grant No.~E429A6M1. X. Li was supported by INFN, sezione di Genova.

\bibliographystyle{apsrev}
\bibliography{reference}

\end{document}